\documentclass[draft]{article}
 
\usepackage{amsmath}
\usepackage{amsthm}
\usepackage{amssymb}
\usepackage{amscd}
\usepackage{epsf}
\usepackage{rotating}
\usepackage{wrapfig}

\chardef\bslash=`\\ 

\makeatletter
\def\verbatim{\interlinepenalty\@M \@verbatim
  \leftskip\@totalleftmargin\advance\leftskip2pc
  \frenchspacing\@vobeyspaces \@xverbatim}
\makeatother
\theoremstyle{plain}

\theoremstyle{remark}

\numberwithin{equation}{section}

\def\1I{\relax{\rm 1\kern-.25em \rm l}} 

\newcommand{\unity}{\1I}

\newcommand{\MyIm}{\Im\mathfrak{m}}
\newcommand{\MyRe}{\Re\mathfrak{e}}

\font\tablefont=cmcsc10

\begin{document}

\thispagestyle{empty}
\rightline{HUB-EP-99/65}
\rightline{hep-th/9912065}
\vspace{2truecm}
\centerline{\bf \Large Supersymmetric M5-branes with $H$-field}
\vspace{.5truecm}

\vspace{1.5truecm}
\centerline{\bf Dieter L\"ust\footnote{luest@physik.hu-berlin.de} and 
                Andr\'e Miemiec\footnote{miemiec@physik.hu-berlin.de}}

\vspace{.4truecm}
{\em 
\centerline{Humboldt-Universit\"at, Institut f\"ur Physik,
D-10115 Berlin, Germany}}

\vspace{1.0truecm}
\begin{abstract}
 In this paper we investigate the form of calibrated M5-branes
 in the presence of a nonvanishing 3-form field H. We discuss the 
 influence of the H-field on the deformation of supersymmetric 
 n-cycles (in particular SLAG submanifolds in ${\mathbb R}^n$).
 In addition we argue for a construction which relates calibrated 
 M5-branes of different ``curved'' dimensions to each other. 
\end{abstract}
\bigskip \bigskip
\newpage


\section{Introduction}

Supersymmetric BPS brane solutions play a very important role in the 
description of string theory and gauge theories. Especially it became 
clear, that one can study gauge theories via various supersymmetric 
brane configurations embedded in 
flat spacetime. This approach is particularly attractive since a lot of 
informations about the nonperturbative gauge dynamics can be obtained 
by lifting the brane configurations to 11d M-theory \cite{witten}, where 
branes preserving a certain amount of the 32 supercharges of M-theory are 
described by calibrated surfaces \cite{BBS,GibPap98,GaLaWe98}.
For example, gauge theories with eight unbroken supercharges ($N\,=\,2$
supersymmetry in four dimensions) correspond to $SU(2)$ Special 
Lagrangian (SLAG) calibrations \cite{witten}.
Similarly, gauge theories with four unbroken supercharges can be
associated to $SU(3)$-SLAG calibrations, the case which was 
discussed in \cite{KLM}.
However, unlike the supersymmetric 2-cycles, the supersymmetric
3-cycles and all higher odd dimensional SLAG cycles are very difficult 
to construct explicitly.
Concrete applications requires a simple handling of such M5-brane 
solutions. Unfortunately for the $SU(3)$-SLAG only extremely special
solutions could be written down so far. Therefore one is looking 
for constructions which connect geometries of different dimensions to each 
other. In this context a very exciting construction exists 
\cite{GeometricAsymptotics},  \linebreak 
\parbox{5cm}{\vspace{0.2cm}
  \refstepcounter{figure}
  \label{figureprojection}
  \begin{center}
  \makebox[5cm]{
     \epsfxsize=5cm
     \epsfysize=4cm
     \epsfbox{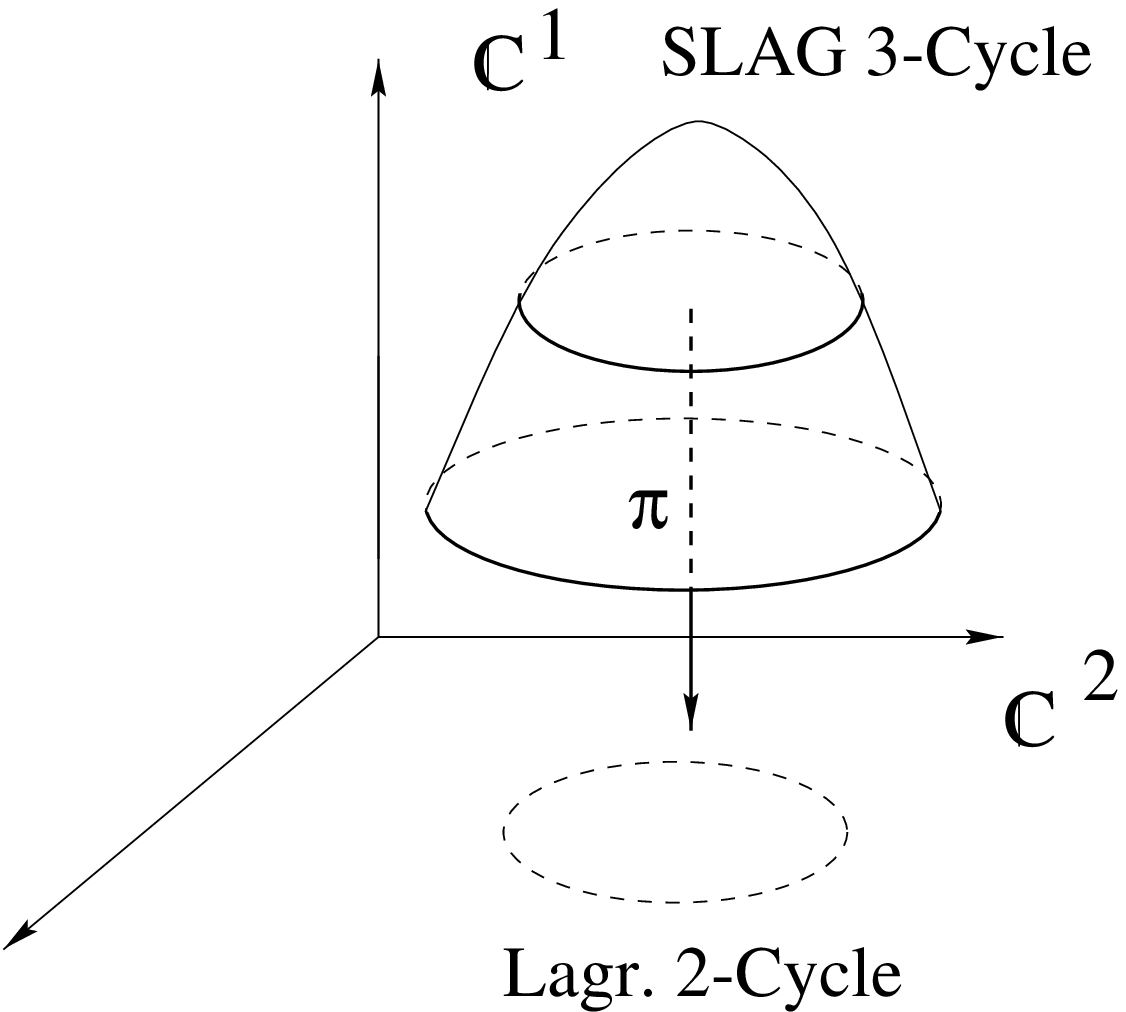}
  }
  \end{center}
  \center{{\bf Fig.{\thefigure}.} 
           The projection principle}
}\par
  \vspace{-5.4cm}
  \hangafter=-14
  \hangindent=5.5cm
  \noindent
which connects Lagrangian submanifolds of different dimensions. 
The construction given there generates a smoothly varying phase 
$\alpha$ on the projected cycles which prevents it to be SLAG 
($\alpha ={\rm const}$) again (see Fig.\ref{figureprojection}). 
We will show that supersymmetric solutions of the latter type exist 
where the non constant phase can be related to the 3 form field $H$ 
which appears in the general equations of motions of the M5-brane\footnote 
{In a recent paper \cite{Strominger} similar modified geometries 
 are investigated in IIA \& IIB theory in the presence of $F$ and $B$ 
 fields.}.
In order to understand this ``projection principle'' we will analyze 
the complete BPS equations of the \hbox{M5-brane}, which includes  
additional terms with the capability to modify the differential 
equations of the SLAG conditions. Such terms appear naturally after turning 
on the 3 form field $H$ and after breaking further 
supersymmetries.\\  
As a byproduct one can use these equations to discuss supersymmetric 
M5-brane solutions in a background including a constant 3 form field $C$ 
of 11d SUGRA. This discussion is also motivated by the recent study of 
Yang-Mills theory in a noncommutative geometry \cite{SW}. 
In this way one might obtain nonperturbative informations about 
noncommutative Yang-Mills theories via $H$-deformed SW-curves.\\
For concreteness, we will focus on the cases of supersymmetric 
2-cycles  and show how the SLAG conditions are affected by the 
$H$-fields. The arguments are generic and can be generalized to all 
$SU(n)$-SLAG cases as well.\\
The paper is organized as follows. In the next section we analyze the 
BPS equations of \cite{GaLaWe98,GaLaWe98b} showing that SLAG submanifolds 
become LAG in the presence of the field $H$.\\[-4ex]

\section{The modified Geometries}

\subsection{The BPS Equations}
\label{SectionBPS}

In static gauge the embedding of the M5-brane into flat 11d spacetime 
is realized by a map $f$, which describes the dependence of the 
transverse coordinates $X_{n'}$, $n' = 1,\,{\ldots}\,,\,5$ on the 
brane coordinates $q_m$, $m=0,\,{\ldots}\,,\,5$. Furthermore there 
lives a two-form field $B_{mn}$ 
on the six-dimensional worldvolume of the M5-brane with field 
strength $H=dB$. A nonlinear self duality constraint is realized
on the field $H$, so that the anti self dual part can be computed 
from the self dual one:\\[-5ex] 
\begin{eqnarray*}
     H_{abc}~=~\frac{1}{Q}\left(\,h_{abc}\,+\,2\,(kh)_{abc}\,\right).\\[-3ex]
\end{eqnarray*}
$k_a^b$ and $Q$ are defined by $k_a^b=h_{acd}h^{bcd}$ and $Q=1-\frac{2}{3}tr k^2$. 
The equations of motion of the M5-brane are obtained from the 
superembedding approach \cite{HoSeWe97a}. 
As it is shown there, in the transformation law of the residual spacetime 
supersymmetry a projector $\Gamma$ appears,\\[-3ex] 
\begin{eqnarray*}
    \delta\Theta = (1-\Gamma)\,\epsilon.\label{SUSY}\\[-4ex]
\end{eqnarray*}
From this formula one can find supersymmetry preserving solutions by 
requiring $\delta\Theta\equiv 0$. Since the M5-brane breaks the 
$SO(1,10)$ Lorentz invariance to a residual $SO(1,5)\times SO(5)$ 
subgroup, the formulas are written in a well adapted form. In particular
the 11d gamma matrices $\bar\Gamma_a$ are constructed out of 
$Spin(1,5)$ and $Spin(5)$ gamma matrices by a general property of
Clifford algebras. For the case at hand we obtain:
\begin{eqnarray}\label{Clifford}  
     {\mathcal{C}}\ell({\mathbb{R}}^{1,5}\oplus 
     {\mathbb{R}}^{5})
     &\buildrel \cong \over \longrightarrow&
     {\mathcal{C}}\ell({\mathbb{R}}^{1,5})\otimes 
     {\mathcal{C}}\ell({\mathbb{R}}^5)\\
      j(v\oplus w) &=& v\otimes\unity + \Gamma_7\otimes w.\nonumber
\end{eqnarray}
$\Gamma_7$ is the chirality operator of the algebra 
${\mathcal{C}}\ell({\mathbb{R}}^{1,5})$. Then the explicit coordinate 
expression of the last formula for a flat 11d SUGRA background as derived 
in \cite{GaLaWe98b} reads:
{\small
\begin{eqnarray*}
 \hat\delta\Theta_\beta{}^j ~=~\hspace{80ex}\\
       \frac{-1}{2\det\,e}\epsilon^{\alpha i}
              \left\{ ~                     
                       \partial_a X^{c'}
                         (\gamma^a)_{\alpha\beta}
                              (\gamma_{c'})_i{}^j
                      -\frac{1}{3!}\partial_{a_1} X^{c_1'}
                       \partial_{a_2} X^{c_2'}\partial_{a_3} X^{c_3'}
                       (\gamma^{a_1a_2a_3})_{\alpha\beta}
                       (\gamma_{c_1'c_2'c_3'})_{i}{}^j
              \right.\hspace{3ex}\\
              \left.  +\frac{1}{5!}\partial_{a_1} X^{c_1'}
                       \partial_{a_2} X^{c_2'}\partial_{a_3} X^{c_3'}
                       \partial_{a_4} X^{c_4'}\partial_{a_5} X^{c_5'}
                       (\gamma^{a_1a_2a_3a_4a_5})_{\alpha\beta}
                       (\gamma_{c_1'c_2'c_3'c_4'c_5'})_{i}{}^j
              \right\}\hspace{3ex}\\
            -\frac{1}{2}\epsilon^{\alpha i}
              \left\{
                      -h^{m_1m_2m_3}\partial_{m_2} X^{c_2'}
                       \partial_{m_3} X^{c_3'}
                       (\gamma_{m_1})_{\alpha\beta}
                       (\gamma_{c_2'c_3'})_i{}^j
                      -\frac{1}{3}h^{m_1m_2m_3}
                       (\gamma_{m_1m_2m_3})_{\alpha\beta}\delta_i{}^j
              \right\}.\hspace{3ex}
\end{eqnarray*}
\begin{eqnarray}\label{BPS}
\end{eqnarray}
}\\[-2ex]
Here $\gamma^i$ are the chiral $Spin(1,5)$ tangent space gamma matrices 
on the M5-brane and $\gamma'{}^i$ the $Spin(5)$ gamma matrices of the 
transverse space.
Without further specialisations the only solution is given by a flat M5-brane
preserving one half of spacetime supersymmetry. On the other hand if there 
are some nontrivial identities of products of gamma matrices applied to the
spinor $\epsilon$ one obtains differential equations which determine 
more general supersymmetric M5-brane solution.   
The amount of supersymmetry preserved by such a solution is determined 
by the number of relations imposed on the 11d gamma algebra.\\  

\subsection{The $\bf SU(n)$-SLAG calibrations}

In the following we concentrate on those projectors which determine 
the Special Lagrangian geometries. They corresponds to  the pattern 
of branes below (the bar denotes a negative sign of the eigenvalues 
of supersymmetries preserved by the brane),\\
\parbox{12cm}
{\hspace{-0.5cm}
 \parbox{6cm}
 {
  \refstepcounter{table}
  \label{SU2SLAG}
  \begin{center}
  \begin{tabular}{|c|ccccccc|}
  \hline
                   & 1 & 2 &   &   &   & 1'& 2'\\
  \hline
  \hline
                M5 & 1 & 2 & 3 & 4 & 5 &   &   \\
$\overline{\rm M5}$&   &   & 3 & 4 & 5 & 6 & 7 \\
  \hline
  \end{tabular} 
  \end{center}
  \center{{\tablefont Table {\thetable}.} \hbox{$SU(2)$-SLAG} }
 }\hspace{2ex}
 \parbox{6cm}
 {
  \refstepcounter{table}
  \label{SU3SLAG}
  \begin{center}
  \begin{tabular}{|c|cccccccc|}
  \hline
                       & 1 & 2 & 3 &   &   & 1'& 2'& 3'\\
  \hline\hline
  M5                   & 1 & 2 & 3 & 4 & 5 &   &   &   \\
  $\overline{\rm M5}$  &   &   & 3 & 4 & 5 & 6 & 7 &   \\
  M5                   &   & 2 &   & 4 & 5 & 6 &   & 8 \\
  \hline
  \end{tabular} \end{center}
  \center{{\tablefont Table {\thetable}.} \hbox{$SU(3)$-SLAG} }
 }
}\\[1ex]
which induce projectors of the type 
$\gamma{}^{ij} ~=~ -\gamma'{}^{ij}$. In the absence of the self dual 
3 form $h_{abc}$ these projectors imply the SLAG differential 
equations. The topology of the solution for $SU(n)$-SLAG is 
${\mathbb{R}}^{1,5-n}\times M_n~$ with $M_n$ the SLAG submanifold. 
The first three terms of eq. (\ref{BPS}) were analyzed in \cite{GaLaWe98}.
For these types of geometries the authors found the BPS equations 
which for $SU(2)$-SLAG and $SU(3)$-SLAG read
{\small
\begin{eqnarray*}
\hat\delta\Theta_\beta{}^j = \frac{-1}{2\det\,e}\,\epsilon^{\alpha i} 
    \left\{
            \frac{1}{2}\underbrace{
                                    \left[\vbox{\vspace{2.5ex}}\right.
                                       \partial_a X_{c}-\partial_c X_{a}
                                    \left.\vbox{\vspace{2.5ex}}\right]
                                  }_{
                                      \hbox{\small generic}
                                  }
            \,\gamma^a\gamma'{}^{c}
            +\left[\vbox{\vspace{2.5ex}}\right.
                    \sum_a\partial_a X_{a}-
                    \underbrace{
                                 \det(\partial X)
                               }_{
                                \hbox{\small 0 for $SU(2)$}
                               }
                \left.\vbox{\vspace{2.5ex}}\right]
             \gamma^{1}\gamma'{}^{1}
    \right\}\\[-4ex]
\end{eqnarray*}
\begin{eqnarray}\label{BPSSLAG}
\end{eqnarray}
}\\[-4ex]
The term in front of $\gamma^{a}\gamma'{}^c$ enforces the 
M5-brane to be a Lagrangian submanifold. That is a submanifold of half 
of the dimension of the embedding space on which the symplectic forms 
restricts to zero. The symplectic form is defined by 
$\omega(X_{\mathbb{R}},Y_{\mathbb{R}})~=~g(JX_{\mathbb{R}},Y_{\mathbb{R}})$
with $J$ the complex structure and $g$ the Euclidean scalar product of the 
embedding space. Using the complex structure one can make a real vector 
$X_{\mathbb{R}}$ into a complex one $X_{\mathbb{C}}$. Since
\begin{eqnarray*}
     <X_{\mathbb{C}},Y_{\mathbb{C}}>_{\mathbb{C}}
  ~=~ g(X_{\mathbb{R}},Y_{\mathbb{R}})
   +i\cdot \underbrace{\omega_{\mathbb{R}}(X_{\mathbb{R}},Y_{\mathbb{R}})}_{0}
\end{eqnarray*}
orthogonality in the real sense implies orthogonality in the complex sense
and vice versa. Then for a manifold to be Lagrangian the complexified 
tangent vectors must be complex orthonormalized. 
Locally the evaluation of the form 
\begin{eqnarray*}
  \Phi ~=~ (\tilde{e}_1+i\tilde{e}_2)\wedge\ldots
           \wedge (\tilde{e}_{2n-1}+i\tilde{e}_{2n}) 
  \label{Calibration}
\end{eqnarray*}  
on the tangent planes ($\tilde{e}_i$ a dual basis of $e_i$, 
$Je_i=e_{i+1}$) measures the phase of the unitary matrix build out 
of these unitary tangent vectors.  
The second term in formula 
(\ref{BPSSLAG}) can be understood as the condition 
$f^\ast\MyIm\,\Phi ~=~0$.\\     
Whereas the geometry of the $SU(2)$-SLAG calibration is nearly trivial, 
the $SU(3)$ generalization is highly complicated. To get a handle for the 
complicated case which are relevant for describing $N=1$ chiral gauge 
theories \cite{KLM} one is looking for constructive methods connecting 
BPS solutions of different dimensions to each other. As explained in  
\cite{GeometricAsymptotics} the simplest construction one can perform on a 
SLAG manifold only preserves the property of being a Lagrangian 
submanifold. Therefore we look for BPS solutions which are Lagrangian but do 
have a smoothly varying phase $\alpha$, which is defined by
$f^\ast\Phi~=~e^{i\alpha}\sqrt{g}\;dq^1\wedge dq^2$ which leads to 
the formula
\begin{eqnarray}\label{tanalpha}
         \tan\alpha &=& \frac{f^\ast\MyIm\,\Phi}{f^\ast\MyRe\,\Phi}.
\end{eqnarray}
The constraints following from supersymmetry are so restrictive 
that a non constant phase $\alpha$ can exist only if we turn on the field 
$h_{abc}$.\\   
To indicate the origin of the interesting terms we consider the contribution
of the 4th and 5th term of eq. (\ref{BPS}) to the BPS equations. These 
terms contain the dependence on the field $h_{abc}$. 
Since $h_{abc}$ is self dual the purely space like components can
be computed from $h_{0ij}$. With \hbox{$\epsilon^{0\ldots5}\,=\,-1$} this 
yields
\begin{eqnarray*}
  \hspace{10ex}
  h^{def}~=~\frac{1}{3!}\epsilon^{defrst}h_{rst} 
         ~=~\frac{1}{2} \underbrace{\epsilon^{def0st}}_{\rm Minkowskian}
            h_{0st}
         ~=~\frac{1}{2} \underbrace{\epsilon^{defst}}_{\rm Euclidean}h_{0st}
\end{eqnarray*}
which motivates the convenient abbreviations 
\begin{eqnarray*}
    F_{ab} &=& h_{0ab}\; \in\;  \Lambda^2({\mathbb{R}}^5),\\
   (\ast F)^{abc} &=& \frac{1}{2}\epsilon^{abcde}h_{0de}\;\in\;
                      \Lambda^3({\mathbb{R}}^5).
\end{eqnarray*}
With these definitions the fourth term can be rewritten like:
{\small
\begin{eqnarray*}
  \hspace{7ex}
  {\rm 4th} &=& -h^{m_1m_2m_3}\partial_{m_2}X^{c_2'}\partial_{m_3}X^{c_3'}
          (\gamma_{m_1})_{\alpha\beta}(\gamma_{c_2'c_3'})_i{}^j\\
      &=& -h^{0ab}\partial_{a}X_{c_2}\partial_{b}X_{c_3}
          (\gamma_{0})_{\alpha\beta}(\gamma'{}^{c_2c_3})_i{}^j
          -h^{def}\partial_{e}X_{c_2}\partial_{f}X_{c_3}
          (\gamma_{d})_{\alpha\beta}(\gamma'{}^{c_2c_3})_i{}^j\\
      &=& F^{ab}\partial_{a}X_{c_2}\partial_{b}X_{c_3}
          \,\gamma_0\,\gamma'{}^{c_2c_3}
          \,-\,(\ast F)^{abc}\partial_{a}X_{c_2}\partial_{b}X_{c_3}
          \,\gamma_{c}\,\gamma'{}^{c_2c_3},
\end{eqnarray*}
}\\[-2ex]
and the last one simplifies to 
{\small
\begin{eqnarray*}
  {\rm 5th} &=& -\frac{1}{3}h^{m_1m_2m_3}(\gamma_{m_1m_2m_3})_{\alpha\beta}
      ~=~ 2F^{ab}\gamma_{0ab}.\hspace{10ex}
\end{eqnarray*}
}\\[-2ex]
Then the BPS equation gets modified in the following way: 
{\small
\begin{eqnarray*}
\hat\delta\Theta_\beta{}^j = \ldots
    -\frac{1}{2}\epsilon^{\alpha i} 
    \left\{
          F^{ab}\partial_{a}X_{c_2}\partial_{b}X_{c_3}
          \,\gamma_0\,\gamma'{}^{c_2c_3}
          \,-\,(\ast F)^{abc}\partial_{a}X_{c_2}\partial_{b}X_{c_3}
          \,\gamma_{c}\,\gamma'{}^{c_2c_3}
          \,+\,2F^{ab}\gamma_{0ab}
    \right\}
\end{eqnarray*}
}\\[-2ex]
If one does not introduce additional projectors, which further reduce 
the amount of supersymmetry, the last equation states that the 
cycle remains $SU(n)$-SLAG but will be deformed by the field $h_{abc}$. 
Technically speaking the moduli of the cycle become functions of the 
field $h_{abc}$  and are partially eliminated in favour of the 
degrees of freedom of the field $h_{abc}$. One the other hand breaking 
SUSY further by additional projectors the $SU(n)$-SLAG conditions 
will be affected as we will discuss now. 

\subsection{Investigation of the basic example}

Now we specialize all considerations for the $SU(2)$-SLAG example 
as given in  {\tablefont Table} \ref{SU2SLAG} which contains the 
generic behaviour. 
One can use a duality relation of the chiral 6d gamma matrices, 
\begin{eqnarray}
   \gamma^{a_1\ldots a_n} ~=~ -\frac{1}{(6-n)!}(-1)^{\frac{n(n+1)}{2}}
            \epsilon^{a_1{\ldots}a_{n}a_{n+1}{\ldots}a_6}
            \gamma_{a_{n+1}\ldots a_{6}},\label{6dduality}
\end{eqnarray}
to further simplify this expression. We want to put as many as possible 
$\gamma$-terms into the form $\gamma_0\gamma'{}^{c_1c_2}$. 
Writing out the corresponding expression for the \hbox{2-cycle} one 
obtains
{\small
\begin{eqnarray*}
 (\ast F)^{abc}\partial_{a}X_{c_2}\partial_{b}X_{c_3}
          \,\gamma_{c}\,\gamma'{}^{c_2c_3}
     &=& 2(\ast F)^{12c}\partial_{1}X_{[1}\partial_{2}X_{2]}
          \,\gamma_{c}\,\gamma'{}^{12}
\end{eqnarray*}
}\\[-2ex]
This has to be discussed for $c=3\,,\ldots\,,\,5$. With the help of the 
projectors and the 6d duality relation eq. (\ref{6dduality}) we may compute,
\begin{eqnarray*}
 \gamma_3\,\gamma'{}^{12}\,\epsilon ~=~ -\gamma_0\,\gamma'{}^{12}\,\epsilon{,}
\end{eqnarray*} 
and $F^{123}~=~F^{45}$. Similarly we may conclude 
$\gamma_{045}~=~-\gamma_{0}\gamma'{}^{12}$.
Then the BPS equations read
{\small
\begin{eqnarray}
\hat\delta\Theta_\beta{}^j=-\frac{1}{2}\det(e^{-1})\epsilon^{\alpha i} 
    \left\{
            \frac{1}{2}\left[\vbox{\vspace{2.5ex}}\right.
                          \partial_a X_{c}-\partial_c X_{a}
                       \left.\vbox{\vspace{2.5ex}}\right]
            \,\gamma^a\gamma'{}^{c}
            +\,\left[\vbox{\vspace{2.5ex}}\right.
                    \sum_a\partial_a X_{a}
                \left.\vbox{\vspace{2.5ex}}\right]
             \gamma^{1}\gamma'{}^{1}
    \right\}\hspace{3ex}\nonumber\\
   -\frac{1}{2}\epsilon^{\alpha i}
      \left\{\vbox{\vspace{4ex}}\right.
        2(F^{12}+F^{45})
         \left(
                \partial_{1}X_{1}\partial_{2}X_{2}
               -\partial_{2}X_{1}\partial_{1}X_{2}
         \right)
         -4(F^{12}+F^{45})
      \left.\vbox{\vspace{4ex}}\right\}\,\gamma_{0}\gamma'{}^{12}\nonumber\\
      +\ldots\hspace{70ex}\label{2Cycle}
\end{eqnarray}
}\\[-2ex]
where the dots are the terms for $c\,=\,4,5$  implying that that
the corresponding components of $F^{ab}$ vanish. As it stands the 
cycle remains $SU(2)$-SLAG and preserves 1/4 of spacetime supersymmetry. 
In fact for this simple case the closure of $H$ implies 
$F^{45}\,=\,-F^{12}\,=\,{\rm const}$. 
A completely different behaviour arises after 
breaking further supersymmetries. Concretely we will break supersymmetry 
by imposing additional projectors which can be translated to the 
following modified brane picture below \cite{GaLaWe98b}:\\ 
\parbox{12cm}
{
 \refstepcounter{table}
 \label{braneconf2c}
 \begin{center}
 \begin{tabular}{|c|ccccccc|}
 \hline
                      & 1 & 2 &   &   &   & 1'& 2'\\
 \hline
 \hline
  M5                  & 1 & 2 & 3 & 4 & 5 &   &   \\
  $\overline{\rm M5}$ &   &   & 3 & 4 & 5 & 6 & 7 \\
  M2                  & 1 &   &   &   &   &   & 7 \\
 \hline
 \end{tabular} 
 \end{center}
 \center{{\tablefont Table {\thetable}.} Brane configuration }
}\\[2ex]
From the additional ``$M2$-brane projector'' 
$\bar\Gamma_{017}\,\epsilon\,=\,\eta\,\epsilon$ ($\bar\Gamma_a$ 11d 
tangent frame gamma matrices, $\eta$ a sign) one obtains
\begin{eqnarray*}
     \bar\Gamma_{067}\,\epsilon &=& -\,\eta\,\bar\Gamma_{16}\,\epsilon
\end{eqnarray*}
and by the construction of $\bar\Gamma_a$ (see eq. \ref{Clifford}) 
this is identical to 
\begin{eqnarray*}
  \gamma_0\gamma'{}^{12}\,\epsilon &=& 
   \eta\,\gamma^{1}\gamma'{}^{1}\,\epsilon{.}
\end{eqnarray*}
This equality generates an inflow from the purely $h$-field terms to 
the term proportional to $\gamma^{1}\gamma'{}^{1}$ and modifies 
the SLAG condition while preserving the Lagrangian 
property\footnote{
           Instead of including the $M2$-brane projector 
           $\bar\Gamma_{017}\,\epsilon~=~\eta\,\epsilon$ 
           one could choose 
           $\bar\Gamma_{016}\,\epsilon~=~\kappa\,\epsilon$, 
           which modifies the Lagrangian condition but 
           preserves the other equation. But this case is not 
           so well suited for our purposes but seems to be 
           the generic one for higher dimensions.}. 
If one writes out the resulting BPS equations, one finds:
{\small
\begin{eqnarray*}
\hat\delta\Theta_\beta{}^j &=&
    -\frac{1}{2}\det(e^{-1})\epsilon^{\alpha i} 
    \left\{
            \frac{1}{2}\left[\vbox{\vspace{2.5ex}}\right.
                          \partial_1 X_{2}-\partial_2 X_{1}
                       \left.\vbox{\vspace{2.5ex}}\right]
            \,\gamma^1\gamma'{}^{2}
            +\,\left[\vbox{\vspace{2.5ex}}\right.
                    \sum_a\partial_a X_{a}
                \left.\vbox{\vspace{2.5ex}}\right]
             \gamma^{1}\gamma'{}^{1}
    \right\}\\
    && -\frac{1}{2}\epsilon^{\alpha i}
      \left\{\vbox{\vspace{4ex}}\right.
        2(F^{12}+F^{45})
         \left(
                \partial_{1}X_{1}\partial_{2}X_{2}
               -\partial_{2}X_{1}\partial_{1}X_{2}
         \right)
         -4(F^{12}+F^{45})
      \left.\vbox{\vspace{4ex}}\right\}\,
      \underbrace{
                   \gamma_{0}\gamma'{}^{12}
                 }_{
                   \eta\gamma^1\gamma'{}^{1}
                 }\\
    &=&
   -\frac{1}{2}\det(e^{-1})\epsilon^{\alpha i} 
    \left\{\vbox{\vspace{4ex}}\right.
            \frac{1}{2}\left[\vbox{\vspace{3ex}}\right.
                          \partial_1 X_{2}-\partial_2 X_{1}
                       \left.\vbox{\vspace{3ex}}\right]
            \,\gamma^1\gamma'{}^{2}
            +\left[\vbox{\vspace{3ex}}\right.
                    \sum_a\partial_a X_{a}+\,\eta\,\cdot\,\det(e)\,\cdot\\
            &&   \left(\vbox{\vspace{2.5ex}}\right.
                      2(F^{12}+F^{45})
                      \left(
                             \partial_{1}X_{1}\partial_{2}X_{2}
                            -\partial_{2}X_{1}\partial_{1}X_{2}
                      \right)
                      -4(F^{12}+F^{45})
                 \left.\vbox{\vspace{2.5ex}}\right)
            \left.\vbox{\vspace{3ex}}\right]\gamma^1\gamma'{}^{1}
     \left.\vbox{\vspace{4ex}}\right\}\\
\end{eqnarray*}
}\\[-3ex]
We can solve for the $F_{ab}$:
\begin{eqnarray}
   F^{12}+F^{45} \,=\,  \frac{\eta}{2\det(e)}\;
                      \frac{
                             \sum_a\partial_a X_{a}
                           }
                           {2-
                            \left(
                               \partial_{1}X_{1}\partial_{2}X_{2}
                              -\partial_{2}X_{1}\partial_{1}X_{2}
                            \right)
                           }\label{BPSH}
\end{eqnarray}
In addition to this constraints one also has to ensure the closure of the 
spacetime 3 form $H_{ijk}~=~ e_i^ae_j^be_k^c\,H_{abc}$. 
The simplest way to satisfy the closure of $H_{ijk}$ is to take the 
components $F^{12}$ and $F^{45}$ to be constant.\\ 
Before we discuss the consequence of this choice let us remind some  
of the geometrical formulas. 
The pullbacks of the embedding space differential forms $\omega$, 
$\MyRe\,\Phi$ and  $\MyIm\,\Phi$ under the map $f$ read
\begin{eqnarray*}
    f^\ast\omega     &=& \left(
                                  \partial_2 X_1
                                 -\partial_1 X_2
                         \right)\,
                         dq^1\wedge dq^1,\\
   f^\ast\MyRe\,\Phi &=& \left(
                                 1-
                                  \partial_1 X_1
                                  \partial_2 X_2  
                                 +\partial_2 X_1  
                                  \partial_1 X_2
                         \right)\,
                         dq^1\wedge dq^2,\\
   f^\ast\MyIm\,\Phi &=& \left(
                                  \partial_1 X_1                           
                                 +\partial_2 X_2
                         \right)\,
                         dq^1\wedge dq^2.
\end{eqnarray*}
We will need the prefactors in front of the differentials which we 
denote by $[f^\ast\omega]$ and so on. 
The induced metric on the embedded submanifold $M^2$ is given by
\begin{eqnarray*}
      g_{ij} &=& \left(
                       \begin{array}{cc}
                          1+\left(\partial_1 X_1\right)^2
                           +\left(\partial_1 X_2\right)^2&
                           \partial_1 X_1\partial_2 X_1
                          +\partial_1 X_2\partial_2 X_2\\[0.5ex]
                           \partial_1 X_1\partial_2 X_1
                          +\partial_1 X_2\partial_2 X_2 &
                          1+\left(\partial_2 X_1\right)^2
                           +\left(\partial_2 X_2\right)^2
                       \end{array}
                 \right)           
\end{eqnarray*}
and the following algebraic identity 
\begin{eqnarray*}
  \det\,g &=&  \left[f^\ast\MyIm\,\Phi\right]^2
              +\left[f^\ast\MyRe\,\Phi\right]^2
              +\left[f^\ast\omega\right]^2
\end{eqnarray*}
holds. Combining this identity with the BPS equation $f^\ast\omega\,=\,0$
and the BPS equation (\ref{BPSH}) one computes for the  
phase (eq. \ref{tanalpha}) the expression below,
\begin{eqnarray*}
        \tan\alpha &=& 
                  2\eta\,(F^{12}+F^{45})\,\sqrt{
                                     \frac{
                                            (1+[f^\ast\MyRe\,\Phi])^2
                                          }
                                          {
                                             1-4\,(F^{12}+F^{45})^2\,
                                             (1+[f^\ast\MyRe\,\Phi])^2 
                                          }
                                   },
\end{eqnarray*}
which is generically not constant. Therefore the cycle is $SU(2)$-SLAG
only if the sum of the two moduli $F^{12}$ and $F^{45}$ vanishes. 
Furthermore for non vanishing $F^{12}+F^{45}$ the cycle is not anymore 
SLAG but still a Lagrangian submanifold\footnote
{For the choice of projectors depicted in 
{\tablefont Table} \ref{braneconf2c}.}. 


\section{Conclusions}

We have shown that for the construction of supersymmetric 
solutions the field $h_{abc}$ can be used in two alternative 
directions. Turning on the field $h_{abc}$ without further breaking 
of supersymmetries does not change the class of calibrated 
submanifolds one started with but only deforms the cycle inside 
this class.
Alternatively one can break additional supersymmetries through 
further projectors, which would led to supersymmetric solutions 
which are not covered by the list of standard calibrations. Here 
we studied the example on an 1/8 supersymmetric 2-cycle which 
is still Lagrangian.   
Following the construction given in \cite{GeometricAsymptotics}
we conjecture that the Lagrangian 2-cycle can be obtained by projecting 
a $SU(3)$-SLAG cycle into a four dimensional subspace. 
Partial evidence for this conjecture was given in the paper.\\ 
\noindent
The limitations of our calculations are mainly concerned with the 
restriction to a flat metric SUGRA background but this can be 
improved in principle. 
Partial results are available for nonflat backgrounds 
\cite{GuPaTo99,Cederwall,Strominger}. These approaches do have close 
relations but are difficult to adapt.


\vskip0.5cm
\noindent{\bf Acknowledgements:}

\noindent 
 We want to thank A.Karch, A.Krause and D.Smith for useful discussions.


\end{document}